\documentclass[aps,prd,reprint]{revtex4-1}

\usepackage[utf8]{inputenc}
\usepackage[T1]{fontenc}

\usepackage{lmodern}
\usepackage{microtype}
\usepackage{amsmath,amsfonts,amssymb,amsthm}
\usepackage{mathrsfs}
\usepackage[hidelinks]{hyperref}
\usepackage{cleveref}
\usepackage{graphicx}
\usepackage{tensor}
\usepackage{braket}
\usepackage{mleftright} \mleftright
\usepackage{comment}
\usepackage{array}
\usepackage{booktabs}
\usepackage{todonotes}
\usepackage{mathtools} % for mathclap

\crefname{equation}{}{}

\newcommand*{\dd}{\mathop{}\!d}

\newcommand*{\pd}{\mathop{}\!\partial}
\newcommand*{\cd}{\mathop{}\!\nabla}
\newcommand*{\vd}{\mathop{}\!\delta}

\newcommand*{\scri}{\ensuremath{\mathscr{I}}}
\newcommand*{\lied}{\mathop{}\!\mathcal{L}}

\newcommand*{\R}{{\mathbb{R}}}

\newcommand*{\bdryeq}{\mathrel{\hat=}}

\newcommand*{\sia}{\mu}
\newcommand*{\sib}{\nu}
\newcommand*{\sic}{\sigma}

\begin{document}

\title{Distinct Minkowski Spaces from BMS Supertranslations}
\author{Friedrich Schöller}
\email{schoeller@hep.itp.tuwien.ac.at}
\affiliation{
  Institut für Theoretische Physik, Technische Universität Wien \\ Wiedner Hauptstraße 8--10/136, 1040 Wien, Austria
}

\begin{abstract}
  This work provides a smooth and everywhere well-defined extension of Bondi-Metzner-Sachs (BMS) supertranslations into the bulk of Minkowski space.
  The supertranslations lead to physically distinct spacetimes, all isometric to Minkowski space.
  This construction is in contrast to the often used, non-smooth BMS transformations that appear in a gauge-fixed description of the theory.
\end{abstract}

\maketitle

\section{Introduction}

A field theory is not only defined by its equations of motions but also its boundary conditions.
Boundary conditions may, for example, be required to make an initial value formulation of the theory well posed, to make an action integral meaningful, or to formulate asymptotic conservation laws.
For whatever reason they are needed, when they are introduced something noteworthy can happen: a gauge transformation loses its defining property of being a transformation between two physically equivalent states, and it becomes a physical symmetry, transforming one state into a physically distinct one.
This work does not go into the details of that process but exemplifies one of its incarnations: Bondi-Metzner-Sachs (BMS) transformations in asymptotically flat spacetimes~\cite{Bondi:1962px,Sachs:1962zza}.

Previous calculations performed in a particular coordinate system suggested that a defect arises when acting with supertranslations on Minkowski space~\cite{Compere:2016jwb}.
In order to preserve the coordinate system the supertranslation generators had to be extended in a particular way into the bulk.
The extension of the generators was not smooth and led to apparent defects in the resulting spacetime.
At least since the coordinate-free formulation of BMS transformations~\cite{Geroch:1977jn}, it is evident (see for example~\cite{Guica:2008mu,Barnich:2010eb}) that if one puts emphasis on the covariant description of the theory and does not fix a particular coordinate system, there is much more freedom in extending the generator.
It will be shown that defects can be avoided by choosing a smooth extension of the supertranslation generator into the bulk.

Asymptotically flat spacetimes can be defined as spacetimes admitting a particular structure at null infinity, a structure which is reviewed in \cref{sec:bms}.
The existence of this structure will be our choice of boundary condition.
Before introducing this structure, diffeomorphisms are gauge symmetries.
After the introduction, diffeomorphisms that do not tend to the identity at infinity cease to be gauge transformations and become physical symmetries.
These symmetries form the group of asymptotic symmetries given by the BMS transformations, reviewed also in \cref{sec:bms}.
Since diffeomorphisms that do tend to the identity at infinity are still gauge transformations, a BMS symmetry can be extended arbitrarily into the interior.
In \cref{sec:bms-minkowski} it is explicitly shown that BMS transformations can be extended into the interior of Minkowski space such that they are well defined and smooth everywhere.

\section{Asymptotically Flat Spacetimes}
\label{sec:bms}

Specifying boundary conditions for a field theory is not a straightforward procedure.
We wish to have boundary conditions that are weak enough to encompass all physically relevant solutions while being strong enough to allow us to draw interesting conclusions.
In general relativity there exists a particularly elegant and successful definition of boundary conditions by conformal completion~\cite{Penrose:1962ij}.
Here one attaches a boundary to the spacetime at infinity and demands that there exists a metric which can be extended to the boundary and is related to the physical metric by a conformal transformation.
The boundary inherits some structure, depending on the cosmological constant.
For asymptotically flat spacetimes the diffeomorphisms that leave this structure invariant form the BMS group.

\subsection{Asymptotic Structure}

Consider a physical spacetime $M$ with dimension bigger than two.
We define an unphysical spacetime $\tilde M$ with boundary $\scri$ such that the interior of $\tilde M$ is diffeomorphic to $M$.
The unphysical spacetime is required to have an unphysical metric $\tilde g_{\sia\sib}$ related to the physical metric $g_{\sia\sib}$ by a conformal transformation
\begin{align}
  \tilde g_{\sia\sib}
  &= \Omega^2 g_{\sia\sib} \,,
\end{align}
where $\Omega$ is some smooth function that vanishes at $\scri$.
The normal vector to $\scri$,
\begin{align}
  \tilde n^\sia
  &= \tilde g^{\sia\sib} \cd_\sib \Omega
    \,,
\end{align}
is required to be nowhere vanishing.
The requirement that the unphysical metric is well defined even at the boundary $\scri$ leads to restrictions on the physical metric.
These are our boundary conditions.
For spacetimes with vanishing cosmological constant and strong enough matter falloff conditions, $\scri$ is a null hypersurface.
These spacetimes are called \textbf{asymptotically flat} at null infinity.
Since $\scri$ is a null hypersurface, $\tilde n^\sia$ is null and lies inside $\scri$.
The \textbf{asymptotic structure} is an equivalence class of pairs $(\underline{\tilde g}{}_{\sia\sib}, \tilde n^\sia)$ evaluated at $\scri$, where the underline denotes the pullback to $\scri$.
Two such pairs are equivalent if they are related by a conformal transformation $(\underline{\tilde g}{}_{\sia\sib}, \tilde n^\sia) \sim (\kappa^2 \underline{\tilde g}{}_{\sia\sib}, \kappa^{-1} \tilde n^\sia)$, with $\kappa$ being some smooth, nonvanishing function.
This equivalence relation is necessary since $\Omega$ can be rescaled by an arbitrary function.

In the following, indices of tensors with a tilde are raised and lowered using the unphysical metric $\tilde g_{\sia\sib}$.

\subsection{BMS Transformations}

We define BMS transformations as asymptotic symmetries following Geroch~\cite{Geroch:1977jn}.
A \textbf{BMS transformation} is defined as a diffeomorphism around $\scri$ that keeps the asymptotic structure invariant.
A \textbf{trivial BMS transformation} is a BMS transformation that keeps $\scri$ fixed.
Any BMS transformation can be combined with a trivial one, such that the pair $(\underline{\tilde g}{}_{\sia\sib}, \tilde n^\sia)$ is invariant, not only its conformal equivalence class.
We will do so in what follows.

An \textbf{infinitesimal BMS transformation} is represented by a smooth vector field around $\scri$ that generates BMS transformations.
Since a vector field $\xi^\sia$ acts with the Lie derivative on the physical metric, the infinitesimal change of the metric is
\begin{align}
  \vd_\xi g_{\sia\sib}
  &= \lied_\xi g_{\sia\sib}
    \,.
\end{align}
It follows that the infinitesimal change of the asymptotic structure is given by
\begin{subequations}
\label{eq:bms-generators}
\begin{align}
  \vd_\xi \underline{\tilde g}{}_{\sia\sib}
  &\bdryeq
    \lied_\xi \underline{\tilde g}{}_{\sia\sib} - 2 \Omega^{-1} \xi^\sic \tilde n_\sic \underline{\tilde g}{}_{\sia\sib}
  \\
  \vd_\xi \tilde n^\sia
  &\bdryeq
    \lied_\xi \tilde n^\sia + \Omega^{-1} \xi^\sic \tilde n_\sic \tilde n^\sia
    \,,
\end{align}
\end{subequations}
where $\bdryeq$ denotes equality at $\scri$.
Note that we assumed $\vd_\xi \Omega = 0$ since $\Omega$ is to be regarded as a fixed background field, independent of the metric.
An infinitesimal BMS transformation can now be defined as a vector field $\xi^\sia$ such that the right-hand sides of \cref{eq:bms-generators} vanish at $\scri$.

The supertranslations are a normal subgroup of BMS transformations defined as follows.
A \textbf{supertranslation} is a BMS transformation that is generated by a smooth vector field $\xi^\sia$ of the form
\begin{align}
  \xi^\sia
  &\bdryeq h \tilde n^\sia
    \,,
\end{align}
with some smooth function $h$.
Because $\xi^\sia$ is smooth it follows that
\begin{align}
  \xi^\sia
  &= h \tilde n^\sia + \Omega w^\sia
    \label{eq:supertranslation-with-w}
    \,,
\end{align}
for some smooth vector field $w^\sia$.
The conditions for $\xi^\sia$ in \cref{eq:supertranslation-with-w} to be an infinitesimal BMS transformation are
\begin{subequations}
\label{eq:supertranslation-bms-condition}
\begin{align}
  \tilde n^\sia \cd_\sia h
  &\bdryeq h f / 2
  \\
  w^\sia \tilde n_\sia
  &\bdryeq - h f / 2
    \,,
\end{align}
\end{subequations}
where $f$ is defined as the limit of $f = \Omega^{-1} \tilde n^\sia \tilde n_\sia$ when approaching $\scri$.
In four spacetime dimensions, if $\scri$ has the usual topology $\mathbb{S}^2 \times \R$, the quotient of the BMS transformations by the supertranslations gives the Lorentz group.
If we require the BMS transformations to be defined only locally, the quotient is much bigger and consists of so-called superrotations~\cite{Barnich:2009se}.

\section{BMS Transformations of Minkowski Space}
\label{sec:bms-minkowski}

We now turn to the question of how a vector field generating BMS transformations can be extended into the bulk of Minkowski space.
We first study the case of supertranslations.

\subsection{Supertranslations}
\label{sec:supertranslations-minkowski}

Consider $n$-dimensional Minkowski space in spherical coordinates with ``retarded time'' coordinate $u = t - r$
and inverse radial distance $y = 1/r$ such that the physical metric reads
\begin{align}
  g_{\sia\sib} \dd x^\sia \dd x^\sib
  &= - \dd u^2 + 2 y^{-2} \dd u \dd y + y^{-2} \dd \Omega_{S_{n-2}}^2
    \,.
\end{align}
We set the conformal factor $\Omega = y$.
The conditions \cref{eq:supertranslation-bms-condition} that $\xi^\sia$ has to obey to be a supertranslation generator are now
\begin{align}
  \pd_u h
  &\bdryeq 0
  \\
  w^y
  &\bdryeq 0
    \,.
\end{align}
We can write equivalently
\begin{align}
  \xi
  &= (T(x^A) + O(y)) \pd_u
    + O(y^2) \pd_y
    + O(y) \pd_A
    \,,
\end{align}
with some smooth function $T$ depending on coordinates $x^A$ of the $(n-2)$-sphere.

\begin{figure}
  \centering
  \includegraphics{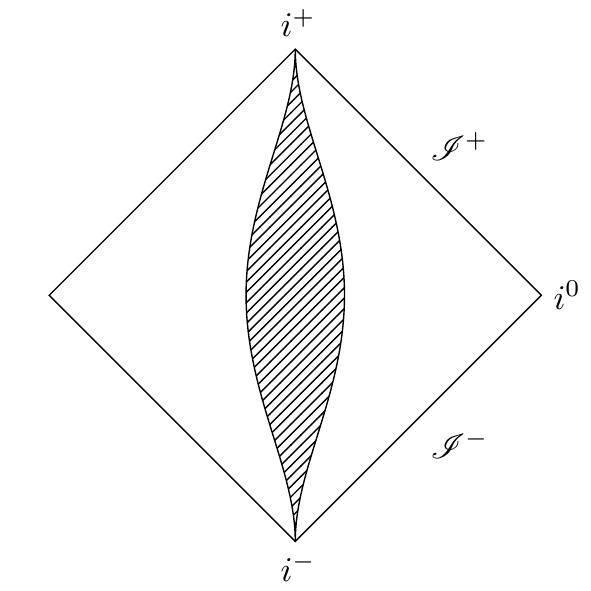}
  \caption{The vector field $\xi^\sia$ is constructed to vanish at the shaded region in Minkowski space.}
  \label{fig:vanishing-xi}
\end{figure}

Now we extend $\xi^\sia$ into the bulk. Since the difference between any two such extensions vanishes at $\scri$, all extensions of the same $\xi^\sia$ are in the same BMS equivalence class. We are free to extend $\xi^\sia$ however we see fit. A convenient choice is
\begin{align}
  \xi
  &= T(x^A) s(r) \pd_t
    \label{eq:xi}
    \,,
\end{align}
given in the coordinate system $(t, r, x^A)$.
Here $s$ is some smooth cutoff function on $\R$, satisfying
\begin{align}
  s(r)
  &=
    \begin{cases}
      0 & r < 1
      \\
      1 & r > 2
      \,,
    \end{cases}
\end{align}
and interpolating in an arbitrary but smooth way between 0 and 1 for $1 \le r \le 2$.
By construction, the vector field $\xi^\sia$ vanishes in some neighborhood around the line $r = 0$ (see \cref{fig:vanishing-xi}).
Therefore, the non-smoothness of $T(x^A)$ at $r = 0$, stemming from the non-smoothness of the coordinates $x^A$ there, is irrelevant.
Since $\pd_t$ is smooth, we conclude that $\xi^\sia$ is smooth everywhere in the bulk.
Since $\xi^\sia$ is time independent, it is simple to integrate, leading to the supertranslation given in coordinates by
\begin{align}
  t'
  &= t + T(x^A) s(r)
    \,.
\end{align}
Supertranslations can therefore be extended to globally well-defined diffeomorphisms.

% \subsection{Independent Supertranslations on \texorpdfstring{$\scri^+$}{I+} and \texorpdfstring{$\scri^-$}{I-}}
% \label{sec:unlock}

% In the previous section, the action of the vector field $\xi^\sia$ on $\scri^+$ was determined by its action on $\scri^-$ and vice versa.
% Since this clearly does not correspond to the most general case, it is of interest to construct a vector field that acts on those two regions independently.
% To do this we use as similar vector field as before, but with an additional factor:
% \begin{align}
%   \xi
%   &= T(x^A) s(r) s(|t|) \pd_t
%     \,.
% \end{align}
% All the essential features of the previous case still hold, but the vector field now vanishes in a neighborhood around the hypersurface $t = 0$.
% It follows that $T$ can be chosen independently for $t > 0$ and $t < 0$, so that the actions on $\scri^+$ and $\scri^-$ are independent as well.

\subsection{Lorentz Transformations}

The group of Poincaré transformations is a semidirect product between the Lorentz group and the translations.
Similarly, the group of BMS transformations is a semidirect product between the Lorentz group and the supertranslations.
In the Poincaré case there is not a single Lorentz subgroup, but there are many, one for each choice of base point around which to rotate or boost.
The different Lorentz subgroups are all related by translations.
The BMS case is similar: there is no unique Lorentz subgroup, but there are many, each one related to another by a supertranslation~\cite{Sachs:1962zza}.

If we pick a Lorentz subgroup of the group of BMS transformations, we can write any BMS transformation as the product of an element of this Lorentz subgroup and a supertranslation.
On Minkowski spacetime there is a preferred choice to pick a Lorentz subgroup by requiring the transformations to be global isometries.
When we pick such a Lorentz subgroup, the combination of a Lorentz transformation and a supertranslation is also globally well defined.
Since this gives us the whole group of BMS transformations, we conclude that BMS transformations are globally well-defined diffeomorphisms when acting on Minkowski space.

\hfill\\
\section{Conclusion}

It was shown that BMS transformations act as smooth diffeomorphisms on Minkowski space.
Each element of the orbit of BMS transformations acting on Minkowski space is therefore isometric to Minkowski space.
The different elements of this orbit can be regarded as different gravitational vacua~\cite{Ashtekar:1981hw,Strominger:2013jfa}.
Since they are all isometric to each other, they are locally indistinguishable from one another.
Asymptotically, however, they have different superrotation charges in three~\cite{Barnich:2010eb} and four~\cite{Barnich:2011mi} dimensions.
% To see how isometric spacetimes can have different charges, it is useful to switch from an active viewpoint of BMS transformations, as employed before, to a passive one.
% In the passive viewpoint, one does not transport the metric with the diffeomorphism, but instead the BMS generator related to the charge, and the location the charge is evaluated at.
From this we see that one should not expect to find sources for superrotation charges localized anywhere in spacetime.
This is consistent with the fact that, in a covariant formulation, charges can be defined only asymptotically, since there are no nontrivial, conserved $n-2$ forms~\cite{Barnich:1994db}.

\begin{acknowledgments}
\uchyph=0
The author thanks Hernán González, Daniel Grumiller, Maria Irakleidou, Behnoush Khavari, Wout Merbis, Stefan Prohazka, Jakob Salzer, and Raphaela Wutte for helpful discussions.
\end{acknowledgments}

\bibliographystyle{utphys}
\bibliography{global}

\providecommand{\href}[2]{#2}\begingroup\raggedright\begin{thebibliography}{10}

\bibitem{Bondi:1962px}
H.~Bondi, M.~G.~J. van~der Burg, and A.~W.~K. Metzner, ``{Gravitational waves
  in general relativity. 7. Waves from axisymmetric isolated systems},''
\href{http://dx.doi.org/10.1098/rspa.1962.0161}{{\em Proc. Roy. Soc. Lond.}
  {\bfseries A269} (1962) 21--52}.
%%CITATION = PRSLA,A269,21;%%.

\bibitem{Sachs:1962zza}
R.~Sachs, ``{Asymptotic symmetries in gravitational theory},''
\href{http://dx.doi.org/10.1103/PhysRev.128.2851}{{\em Phys. Rev.} {\bfseries
  128} (1962) 2851--2864}.
%%CITATION = PHRVA,128,2851;%%.

\bibitem{Compere:2016jwb}
G.~Compère and J.~Long, ``{Vacua of the gravitational field},''
  \href{http://dx.doi.org/10.1007/JHEP07(2016)137}{{\em JHEP} {\bfseries 07}
  (2016) 137},
\href{http://arxiv.org/abs/1601.04958}{{\ttfamily arXiv:1601.04958 [hep-th]}}.
%%CITATION = ARXIV:1601.04958;%%.

\bibitem{Geroch:1977jn}
R.~Geroch, {\em Asymptotic Structure of Space-Time},
  \href{http://dx.doi.org/10.1007/978-1-4684-2343-3_1}{pp.~1--105}.
\newblock Springer US, Boston, MA, 1977.

\bibitem{Guica:2008mu}
M.~Guica, T.~Hartman, W.~Song, and A.~Strominger, ``{The Kerr/CFT
  Correspondence},'' \href{http://dx.doi.org/10.1103/PhysRevD.80.124008}{{\em
  Phys. Rev.} {\bfseries D80} (2009) 124008},
\href{http://arxiv.org/abs/0809.4266}{{\ttfamily arXiv:0809.4266 [hep-th]}}.
%%CITATION = ARXIV:0809.4266;%%.

\bibitem{Barnich:2010eb}
G.~Barnich and C.~Troessaert, ``{Aspects of the BMS/CFT correspondence},''
  \href{http://dx.doi.org/10.1007/JHEP05(2010)062}{{\em JHEP} {\bfseries 05}
  (2010) 062},
\href{http://arxiv.org/abs/1001.1541}{{\ttfamily arXiv:1001.1541 [hep-th]}}.
%%CITATION = ARXIV:1001.1541;%%.

\bibitem{Penrose:1962ij}
R.~Penrose, ``{Asymptotic properties of fields and space-times},''
\href{http://dx.doi.org/10.1103/PhysRevLett.10.66}{{\em Phys. Rev. Lett.}
  {\bfseries 10} (1963) 66--68}.
%%CITATION = PRLTA,10,66;%%.

\bibitem{Barnich:2009se}
G.~Barnich and C.~Troessaert, ``{Symmetries of asymptotically flat 4
  dimensional spacetimes at null infinity revisited},''
  \href{http://dx.doi.org/10.1103/PhysRevLett.105.111103}{{\em Phys. Rev.
  Lett.} {\bfseries 105} (2010) 111103},
\href{http://arxiv.org/abs/0909.2617}{{\ttfamily arXiv:0909.2617 [gr-qc]}}.
%%CITATION = ARXIV:0909.2617;%%.

\bibitem{Ashtekar:1981hw}
A.~Ashtekar, ``{Radiative Degrees of Freedom of the Gravitational Field in
  Exact General Relativity},''
\href{http://dx.doi.org/10.1063/1.525169}{{\em J. Math. Phys.} {\bfseries 22}
  (1981) 2885--2895}.
%%CITATION = JMAPA,22,2885;%%.

\bibitem{Strominger:2013jfa}
A.~Strominger, ``{On BMS Invariance of Gravitational Scattering},''
  \href{http://dx.doi.org/10.1007/JHEP07(2014)152}{{\em JHEP} {\bfseries 07}
  (2014) 152},
\href{http://arxiv.org/abs/1312.2229}{{\ttfamily arXiv:1312.2229 [hep-th]}}.
%%CITATION = ARXIV:1312.2229;%%.

\bibitem{Barnich:2011mi}
G.~Barnich and C.~Troessaert, ``{BMS charge algebra},''
  \href{http://dx.doi.org/10.1007/JHEP12(2011)105}{{\em JHEP} {\bfseries 12}
  (2011) 105},
\href{http://arxiv.org/abs/1106.0213}{{\ttfamily arXiv:1106.0213 [hep-th]}}.
%%CITATION = ARXIV:1106.0213;%%.

\bibitem{Barnich:1994db}
G.~Barnich, F.~Brandt, and M.~Henneaux, ``{Local BRST cohomology in the
  antifield formalism. 1. General theorems},''
  \href{http://dx.doi.org/10.1007/BF02099464}{{\em Commun. Math. Phys.}
  {\bfseries 174} (1995) 57--92},
\href{http://arxiv.org/abs/hep-th/9405109}{{\ttfamily arXiv:hep-th/9405109
  [hep-th]}}.
%%CITATION = HEP-TH/9405109;%%.

\end{thebibliography}\endgroup

\end{document}